\RequirePackage{fix-cm}

\documentclass[smallextended]{svjour3}       
\smartqed  

\usepackage{lineno,hyperref}
\usepackage{graphicx}
\usepackage{graphicx}
\usepackage{subcaption}
\usepackage{soul}
\usepackage{comment}

\usepackage{enumerate}

\usepackage{float}
\usepackage{color}
\mathchardef\mhyphen="2D
\usepackage{comment}
\usepackage{listings}
\usepackage{amsfonts}

\usepackage[utf8]{inputenc}
\usepackage[english]{babel}
\usepackage{fancyhdr}
\usepackage{lastpage}
\usepackage{amsmath}

\usepackage[margin=1in]{geometry}

\mathchardef\mhyphen="2D

\usepackage[margin=1in]{geometry}

\begin{document}



\title{Combined Effects of Fluid Type and Particle Shape on Particles Flow in Microfluidic Platforms}

\titlerunning{Fluid Type and Particle Shape Affect Trajectories of Particles in Microfluidic Platforms} 

\author{Hakan Ba\c sa\u gao\u glu  \and
        Justin Blount \and
        Sauro Succi \and
        Christopher J. Freitas
}

\institute{H. Ba\c sa\u gao\u glu \at Mechanical Engineering Division, Southwest Research Institute, San Antonio, TX 78238 USA; Currently at Edwards Aquifer Authority, San Antonio, TX 78215 \\
              Tel. : +1-210-4775105,
              email: {hbasagaoglu@edwardsaquifer.org} \and
J. Blount \at Defense Intelligence Solutions Division, Southwest Research Institute, San Antonio, TX 78238 USA \and
S. Succi \at  Fondazione Istituto Italiano di Tecnologia, 
Center for Life Nanoscience at la Sapienza, Rome, vle Regina Margherita, 00165-Italy; Istituto Applicazioni del Calcolo,
Via dei Taurini 19, 00185, Roma, Italy \and
C. J. Freitas \at Mechanical Engineering Division, Southwest Research Institute, San Antonio, TX 78238 USA}

\date{Received: date / Accepted: date}

\maketitle

\begin{abstract}
Recent numerical analyses to optimize the design of microfluidic devices for more effective entrapment or segregation of surrogate circulating tumor cells (CTCs) from healthy cells have been reported in the literature without concurrently accommodating the non-Newtonian nature of the body fluid and the non-uniform geometric shapes of the CTCs. Through a series of two-dimensional proof-of-concept simulations with increased levels of complexity (e.g., number of particles, inline obstacles), we investigated the validity of the assumptions of the Newtonian fluid behavior for pseudoplastic fluids and the circular particle shape for different-shaped particles (DSP) in the context of microfluidics-facilitated shape-based segregation of particles. Simulations with a single DSP revealed that even in the absence of internal geometric complexities of a microfluidics channel, the aforementioned assumptions led to 0.11-0.21$W$ ($W$ is the channel length) errors in lateral displacements of DSPs, up to 3-20$\%$ errors in their velocities, and 3-5$\%$ errors in their travel times. When these assumptions were applied in simulations involving multiple DSPs in inertial microfluidics with inline obstacles, errors in the lateral displacements of DSPs were as high as 0.78$W$ and in their travel times up to 23$\%$, which led to different (un)symmetric flow and segregation patterns of DSPs. Thus, the fluid type and particle shape should be included in numerical models and experiments to assess the performance of microfluidics for targeted cell (e.g., CTCs) harvesting.        

\end{abstract}
\section{Introduction}
\label{intro}

Microfluidic devices with distinct geometric peculiarities have been proposed and tested for size-based and/or shape-based segregation of targeted cells in diverse applications. A microfluidic device with a narrow channel connected to an expanded region with multiple outlets was used to sort out \textit{Euglena gracilis}, microalgea explored for biodiesel and biomass production, based on their geometric shapes with different cell aspect ratios \cite{LMSC17}. Similarly, various microfluidics methods and geometric designs \cite{BHLL11,CMW13} have been developed in cancer research to segregate rare circulating tumor cells (CTCs), typically occuring 0–10 CTCs/mL of blood\cite{HV14,GKK18}, from leukocytes in blood samples for more effective, non-invasive diagnosis and prognoses of tumor progression and metastasis \cite{DASM13,HWX18}.        

Among different mechanisms, deterministic lateral displacement (DLD) and inertial focusing (IF) have been   implemented for shape- and/or size-based separation of cells in microfluidics\cite{BMC18}. For DLD, microfluidics typically contain  microsize inline obstacles in particular arrangements to form desired microflow patterns. Through the DLD, small- and large-sized spherical particles were segregated via microfluidics with an array of prism-shaped inline obstacles, in which segregation of particles was governed by distinct migration pathways different-sized particles experienced along the streamlines\cite{HCA04}. Similarly, using an array of I-shaped inline obstacles in a microfludic device, non-spherical cells were isolated from spherical cells, based on differences in their geometric shape-dependent angular momentum \cite{ZRZ13}. 

IF, based on inertial migration of particles, has been extensively used in label-free separation devices for cell segregation \cite{HCK11,NP13,PCD17}. In this method, size- or shape-based segregation of particles are largely governed by competition and dynamic interactions between particle-fluid hydrodynamics, shear gradients, and wall-lift forces. Relative effects of these factors on the particles transport can be adjusted in part by modifying the device geometry to accomplish shape- and/or size-based enrichment of targeted cells.  

Regardless of targeted cell separation mechanisms, geometric design details of a microfluidic device are imperative for size- and/or shaped-based cell sorting. Numerical models can be used to quantify the underlying competing pore-scale processes to optimize the microfluidic device geometry for more effective cell segregation. We recently reported a new numerical model \cite{BSWB18}, formulated based on the lattice Boltzmann model (LBM), to simulate settling or flow of a mixture of two-dimensional (2D) different-shaped particles (DSP) in a Newtonian fluid. Using this new model (DSP-LBM hereafter), we reported non-negligible errors in flow or settling trajectories and velocities of DSPs as well as in their microfluidics-facilitated shape-based segregation, if their actual geometries are approximated by a circular-cylindrical (circular hereafter) shape. 

Although novel simulations with a mixture of settling or flowing DSPs were reported in \cite{BSWB18}, the DSP-LBM was limited to Newtonian fluid flow simulations, which could hamper its use as a numerical tool to optimize microfluidic device designs to isolate targeted non-uniform shaped cells from body fluids. In microfluidics-facilitated CTCs enrichment studies, as an example, blood is a non-Newtonian fluid \cite{YG08,LMM16}, as its apparent viscosity would decrease in microchannels due to the Fahraeus effect \cite{KMJ10}, and CTCs exhibit non-uniform sizes and morphology \cite{HV14,PAD14,MBL14}. Therefore, the non-Newtonian nature of the body fluid and non-uniform morphology of the particles should be addressed in numerical models supporting the design of microfluidics to segregate disease-causing cells. However, recent mesoscale numerical studies \cite{PCD17,HC17,KMFD16,Da13,MSAM12,JSH15,DTF15,SAM18} focusing on size- and shape-based entrapment or segregation of particles from non-uniform suspensions flowing in microfluidics did not \textit{simultaneously} accommodate the non-Newtonian behavior of the fluid flow and non-uniform shapes of particles. Although the fluid was assumed to be Newtonian in assessing the performance of microfluidic cell capture or segregate devices in \cite{PCD17,KMFD16,MSAM12,JSH15,DTF15,SAM18}, potential errors associated with this assumption have not been reported to date.   

Thus, the main motivation of this paper is to assess potential errors in trajectories and velocities of a mixture of non-uniform shaped particles in a pseudoplastic fluid in  microchannels, if the pseudoplastic fluid is approximated by a Newtonian fluid and the geometric shape of the particles is assumed to be circular. This assessment is crucial as the performance of microfluidic cell capture or segregation devices has been commonly tested using a Newtonian fluid as in \cite{PCD17,HC17,KMFD16,MSAM12,JSH15,DTF15,SAM18}. The secondary motivation is to present a new numerical model for simulating flow of a mixture of DSPs in a non-Newtonian fluid in microfluidics with complex geometrical features, which is suitable for simulating fate and transport of CTCs in body fluid. Unlike the particles flow model that simulates particles as thin solid shells filled with a viscous fluid \cite{MSAM12}, the DSP-LBM simulates particles with intra-particle non-viscous `ghost' fluid that does not contribute to particle-fluid hydrodynamics. Therefore, the DSP-LBM is readily suitable for simulating high-frequency rotations of settling or flowing multiple non-uniform shaped particles, including both discretized curve-shaped and angular-shaped particles  \cite{BSWB18}.

In this paper, we report for the first time the DSP-LBM simulations of a \textit{mixture} of DSPs in non-Newtonian fluid flow in a microflow channel with an array of inline obstacles. Because CTCs are less deformable than white blood cells \cite{DTF15}, DSP-LBM simulations focused on the behavior of mixture of rigid, non-uniform shaped particles in  non-Newtonian fluid flow in microchannels. 2D DSP-LBM simulation with Newtonian and non-Newtonian fluids were performed and compared to investigate the effect of (i) non-Newtonian fluid behavior (pseudoplastic or dilatant) on the lateral displacements of an individual DSP in a microchannel; (ii) flow strength, inertial focusing, fluid type (pseudoplastic vs. Newtonian), and particle shape on the travel times and flow trajectories of a mixture of DSPs in a microchannel without inline obstacles; and (iii) the order of equally-spaced DSPs released from multiple-ports near the inlet on the flow trajectories of DSPs in a Newtonian or pseudoplastic fluid in inertial microfluidics with I-shaped inline obstacles. DSP-LBM simulations demonstrated that different geometric shapes of the particles (i.e., surrogate cells) and the non-Newtonian behavior of the fluid should be accommodated in microfluidic experiments and numerical simulations to assess or optimize the performance microfluidic device geometries for enhanced size-based and shape-based cell enrichment from body fluids. 

\section{Lattice-Boltzmann Model (LBM) for DSPs in Non-Newtonian Fluid Flow}
\label{sec:LBM}
In the LBM \cite{HS89,BSV92,S01,W00}, the mesodynamics of the incompressible, non-Newtonian fluid flow \cite{GDK05,PKYPK07,HR11,DNKS14} can be described by a single relaxation time (SRT)  \cite{BGK54}

   \begin{equation}
  \label{e.LB1} f_{i}\left(\mathbf{r+e}_{i}{\triangle t},t+{\triangle
  t}\right) -f_{i}\left(\mathbf{r},t \right) =\frac{\triangle t}{\tau^{*}} [
  {f_{i}^{eq}\left(\mathbf{r},t \right)-f_{i}\left(\mathbf{r},t
  \right) } ],
  \end{equation}
 
\noindent where $f_i(\mathbf{r},t)$ is the complete set of population densities of discrete velocities $\mathbf{e}_i$ at position $\bf{r}$ and discrete time $t$ with a time increment of $\triangle t$, $\tau^{*}$ is the relaxation parameter associated with non-Newtonian fluid flow, and $f_i^{eq}$ is the local equilibrium \cite{QDL92}, described by

 \begin{equation}
 \label{e.LB2}f_{i}^{eq}=\omega_i \rho \left
 (1+\frac{\mathbf{e}_i\mathbf{\cdot}\mathbf{u}}{c_s^2}
 +\frac{(\mathbf{e}_i\mathbf{\cdot}\mathbf{u})^2}{2c_s^4}-
 \frac{\mathbf{u \mathbf{\cdot}u}}{2c_s^2}\right).
\end{equation}

\noindent where $\omega_i$ is the weight associated with $\mathbf{e}_i$ and $c_s$ is the speed of sound, $c_s= \triangle x /\sqrt(3) \triangle t$, and the local fluid density, $\rho$, and velocity, $\mathbf{u}$, at the lattice node are given by $\rho=\sum_{i} f_{i}$ and $\rho \mathbf{u}=\sum_{i} f_{i}\mathbf{e}_i+\tau^{*} \rho \mathbf{g}$, where $\mathbf{g}$ is the strength of an external force \cite{BG00} and $\tau^{*}=0.5+3\nu^{*}\left(\triangle t /  \triangle x^2 \right)$. The kinematic viscosity of the non-Newtonian fluid is described as $\nu^{*}= \left[ 2^{n-1}   {| \Pi_{D} |}  ^  {\frac{n-1}{2}} \right]  \xi$ \cite{DNKS14,BHNS17}, in which $\xi$ is the consistency,  $\Pi_{D}$ is the second invariant of the rate of strain tensor, $n$ is the fluid-type identifier, $n<1$, $n=1$, and $n>1$ correspond to pseudoplastic (shear-thinning), Newtonian, and dilatant (shear-thickening) fluids, respectively. $\Pi_{D}$ is computed as

 \begin{equation}
 \label{nn3} 
 \begin{aligned}
 \Pi_D = \frac{1}{2} \left(  \left[ tr\left(\mathbf{D} \right) \right]^2  - tr(\mathbf{D}^2) \right),
  \end{aligned}
 \end{equation}

\noindent where $\mathbf{D}=\frac{1}{2} \left( \nabla \mathbf{u}(\mathbf{x})+  (\nabla \mathbf{u}(\mathbf{x)})^T \right)$, $\mathbf{u}=\left(u,v\right)$, $\mathbf{x}=\left(x,y \right)$, $T$ is the transpose, and $tr$ is the trace of the matrix $\mathbf{D}$.  A D2Q9 (two-dimensional nine velocity vector) lattice \cite{S01} was used in numerical simulations.
$\xi=\nu$ for the Newtonian fluid, in which $\nu$ is the kinematic viscosity of the Newtonian fluid. $\tau^{*}$ is related to $\nu$ via $\tau^{*}=0.5+\left( \tau-0.5 \right) \nu^{*} / \nu$, in which $\tau$ is the relaxation parameter associated with the Newtonian fluid. Through the Chapman-Enskog approach, the LB method for a single-phase non-Newtonian fluid flow recovers the Navier-Stokes equation in the limit of small Knudsen number for weakly compressible fluids  ($\triangle \rho /\rho \sim {M}^2 \sim 1 \times 10^{-4}$, where $M$ is the Mach number), in which $\nabla \cdot \mathbf{u}\sim 0$ and $\partial_{t} \mathbf{u} +\left( \mathbf{u} \cdot \nabla \right) \mathbf{u} = -\frac{\nabla P}{\rho}+ \nu^{*} \nabla^2 \mathbf{u} + \mathbf{g}$. Pressure differential, $P$, is computed via the ideal gas relation, $P=c_s^2 \rho$. 

\vspace{5mm} 

\noindent \textbf{Different-shaped particles}. 2D simulations of flow of particles in Newtonian or non-Newtonian fluids were performed in this paper using the DSP-LBM that accommodates particle-fluid hydrodynamics of DSPs \cite{BSWB18}. Simulations were conducted using discretized angular shaped particles (DAsPs), encompassing rectangular and hexagonal particles, and discretized curved-shaped particles (DCsPs), encompassing circular-cylindrical and elliptical particles. The DSP-LBM calculates first the coordinates of the boundary nodes, $\left(x_i, y_i \right)$, on the circumference of DAsPs, using the information on the center of the mass of a particle $\mathbf{x}_c=\left(x_c, y_c \right)$ and other geometric shape-specific parameters. $\left(x_i, y_i \right)$ for circular and elliptical particles are computed by Eqs. \ref{eq_circular} and \ref{eq_elliptical}, respectively,

 \begin{equation} 
  \label{eq_circular} 
 \left[ \begin{array}{c} x_i\\ y_i \end{array} \right] = \left[ \begin{array}{c} x_c \\ y_c \end{array} \right]+R_p \left[ \begin{array}{c} cos\left( 2\pi \left( i-1 \right) / \left(N_{bnd}-1 \right)  \right) \\ sin \left( 2 \pi \left( i-1 \right)  / \left( N_{bnd}-1 \right)  \right) \end{array} \right], 
 \end{equation}
\begin{equation} 
  \label{eq_elliptical} 
 \left[ \begin{array}{c} x_i\\ y_i \end{array} \right] = \left[ \begin{array}{c} x_c \\ y_c \end{array} \right]+  \begin{bmatrix} cos(\Phi_i) cos(\hat{\alpha}) & - sin (\Phi_i)  sin(\hat{\alpha}) \\ cos(\Phi_i) sin(\hat{\alpha})  &  sin (\Phi_i)  cos(\hat\alpha) \end{bmatrix} \left[ \begin{array}{c} c/2 \\ d/2 \end{array} \right], 
 \end{equation}

\noindent where $R_p$ is the radius of a circular particle, $c$ and $d$ are the length of the major and minor axes of an elliptical particle,  $\hat\alpha$ is the initial tilt angle of an elliptical particle in the clockwise direction, $\Phi_i=2\pi \left( i-1 \right) / {\left( N_{Nbd}-1 \right)}$, and $N_{Bnd}$ is the number of boundary nodes. Different from DCsPs, the DSP-LBM calculates first the coordinates of vertices, $\left(x_v, y_v \right)$, for DAsPs, based on the information on $\left(x_c, y_c \right)$ and  other geometric shape-specific parameters. $\left(x_v, y_v \right)$ for hexagonal and rectangular particles are computed by Eqs. \ref{eq_hexagonal} and \ref{eq_rectangular}, respectively,

 \begin{equation} 
  \label{eq_hexagonal} 
 \left[ \begin{array}{c} x_{v_{Hi}}\\ y_{v_{Hi}} \end{array} \right] =\left[ \begin{array}{c} x_c \\ y_c \end{array} \right]+   L  \left[ \begin{array}{c} cos \left(\alpha+\left(i-1 \right) \pi /3 \right) \\  sin\left(\alpha+\left( i-1 \right) \pi /3\right) \end{array} \right], 
 \end{equation}

\begin{equation} 
  \label{eq_rectangular} 
 \left[ \begin{array}{c} x_{v_{Rj}}\\ y_{v_{Rj}} \end{array} \right] =\left[ \begin{array}{c} x_c \\ y_c \\ \end{array} \right]+  \frac{ \sqrt{l^2+w^2} }{2}  \left[ \begin{array}{c} \varphi_{xj}  \\  \varphi_{yj} \\  \end{array} \right], 
 \end{equation}

\noindent where $i \epsilon \left[1,6\right]$, $L$ is the side length of a hexagonal particle, $j \epsilon \left[1,4\right]$, $l$ and $w$ are the long and short side lengths of a rectangular particle,  $\varphi_{xj}=\left[ cos \left(\alpha +\theta \right),-cos \left(\alpha -\theta \right),-cos \left(\alpha +\theta \right),cos \left(\alpha -\theta \right)  \right]$ and $\varphi_{yj}=\left[sin \left(\alpha +\theta \right),  sin \left(\alpha -\theta \right), sin \left(\alpha -\theta \right), -sin \left(\alpha +\theta \right), -sin \left(\alpha -\theta \right) \right]$, and $\alpha$ is the initial tilt angle of a particle (rectangular or hexagonal) in the counterclockwise direction. $x_c= \frac{1}{N} \sum_{i=1}^N {x_i}$ and $y_c=\frac{1}{N} \sum_{i=1}^N {y_i}$, where $N$ is the number of vertices for DAsPs or the number of boundary nodes for DCsPs.

\vspace{5mm} 

\noindent \textbf{Intra-Particle Boundary Nodes (IPBNs) and Extra-Particle Boundary Nodes (EPBNs)}. 
Each DAsP or DCsP is represented by a polygon in the DSP-LBM. The winding number algorithm\cite{O98} is used to determine the location of lattice nodes inside and closest to the particle surface. These nodes are labeled as IPBNs and denoted by $\mathbf{r_v}$. A lattice node outside the particle surface and separated from the closest IPBN by $\mathbf{e_i}$ is labeled as EPBN and represented by $\left( \mathbf{r_v}+\mathbf{e_i} \right)$. Each $\left( \mathbf{r_v},\mathbf{r_v}+\mathbf{e_i} \right)$ pair forms a hydrodynamic link across the particle surface along which the mobile DAsP or DCsP exchanges momentum with the bulk fluid. Particle-fluid momentum exchanges occur at all boundary nodes located at $\mathbf{r}_b=0.5 \left[ \mathbf{r_v}+\left( \mathbf{r_v}+\mathbf{e_i} \right) \right]$. Particle motion is determined by local velocities, $\mathbf{u}_{\mathbf{r}_b}$, computed from particle-fluid hydrodynamics at each $\mathbf{r}_b$.

\vspace{5mm} 

\noindent  \textbf{Particle-fluid Hydrodynamics}. Following the approach in \cite{L94a,NL02,BS10}, population densities near particle surfaces are modified to account for momentum-conserving particle-fluid collisions. Particle-fluid hydrodynamic forces, $\mathbf{F}_{\mathbf{r}_b}$, at each $\mathbf{u}_{\mathbf{r}_b}$ is computed by 

 \begin{equation}
 \label{e.pfh3} \mathbf{F}_{\mathbf{r}_b}=-2\left[f^{\prime}_i\left(\mathbf{r}_v+\mathbf{e}_i \triangle t,t^{\ast}\right)+
\frac{\rho\omega_i}{c_s^2}\left(\mathbf{u}_{\mathbf{r}_b} \cdot
\mathbf{e}_i\right)\right]\mathbf{e}_i.
 \end{equation}

Eq.\ref{e.pfh3} was derived based on the premise that the fluid occupies the entire flow domain to ensure the continuity in the flow field to avoid large artificial pressure gradients that may arise from the compression and expansion of the fluid near particle surface;  however, the fluid inside the particle does not contribute to $\mathbf{F}_{\mathbf{r}_b}$. The translational velocity, $\mathbf{U}_p$, and the angular velocity of the particle, $\Omega_p$, are computed by $\mathbf{U}_p\left( t+\triangle t\right) \equiv \mathbf{U}_p\left(t\right) + \triangle t \left[ \frac{\mathbf{F_T \left( t \right) }}{{m_p}} +\frac{(\rho_p-\rho)}{\rho_p} \mathbf{g}  \right]$ and 
$\Omega_p\left(t+\triangle t\right) \equiv \Omega_p\left(t\right)+\frac{\triangle t}{I_p}\mathbf{T}_T\left( t\right)$, where $\mathbf{F_T}$ and $\mathbf{T}_T$ are the total hydrodynamic force and torque on the particle exerted by the surrounding fluid, respectively. $m_p$ is the particle mass, $I_p$ is the moment of inertia of the particle (Table \ref{tab:parameters}), and $\mathbf{u}_b=\mathbf{U}_p+\Omega_p\times\left({\mathbf{r}_b} 
- \mathbf{r}_c \right)$. 


\begin{table}[h!]
\caption{Mass and moment of inertia of DSPs.}\label{tab:parameters}
\centering
\begin{tabular}{lcr}
\hline
Particle Shape & Particle Mass per Unit Particle Thickness, $m_p$ & Moment of Inertia, $I_p$\\
\hline
$Circular*  $      &     $\pi R^2 \rho_p$   &          $\left( 1/2 \right) m_p {R}^2$\\
$Elliptical$      &      $\pi \left( cd/4 \right) \rho_p$   &         $\frac{m_p}{16} \left( c^2 + d^2 \right)$\\
$Hexagonal $      &     $\left(3/2\right) \sqrt3 L^2 \rho_p $  & $\left(m_p/24 \right) L^2 \left[ 1+3cot^2 \left(  \pi / 6 \right)  \right]$ \\
$Rectangular$     &     $lw\rho_p$                     & $\frac{m_p}{12} \left( l^2+h^2 \right)$\\
\hline

\end{tabular}

\small{(*) the circular particle is treated as a thin solid disk.}
\end{table}

Local forces and torques associated with particle-fluid hydrodynamics at $\mathbf{r}_b$, covered/uncovered nodes due to the particle motion, and steric (repulsive) interaction between particles in close proximity or particles in close contact with stationary solid objects contribute to $\mathbf{F_T}$ and $\mathbf{T}_T$,

\begin{equation}
 \label{e.5}
 \mathbf{F}_T=\sum_{\mathbf{r}_b}
\mathbf{F}_{\mathbf{r}_b}+\sum_{\mathbf{r}^{c,u}_b}\mathbf{F}_{\mathbf{r}^{c,u}_b}
+\sum_{\mid \mathbf{r}_{pw} \mid \le \mid \mathbf{r}_{it} \mid} \mathbf{F}_{\mathbf{r}_{pw}}
+\sum_{\mid \mathbf{r}_{pp'} \mid \le \mid \mathbf{r}_{it} \mid} \mathbf{F}_{\mathbf{r}_{pp'}},
\end{equation}

\begin{eqnarray}
 \label{e.6}
\mathbf{T}_T &=& \sum_{\mathbf{r}_b} \left( \mathbf{r}_b - \mathbf{r}_c
\right)\times \mathbf{F}_{\mathbf{r}_b}+\sum_{\mathbf{r}^{c,u}_b}
\left( \mathbf{r}^{c,u}_b - \mathbf{r}_c \right) \times
\mathbf{F}_{\mathbf{r}^{c,u}_b}\nonumber
+ \sum_{\mid \mathbf{r_{pw}} \mid \le \mid \mathbf{r}_{it} \mid} \left( \mathbf{r}_w - \mathbf{r}_c \right) \times \mathbf{F}_{\mathbf{r}_{pw}} \\
&+& \sum_{\mid \mathbf{r_{pp'}} \mid \le \mid \mathbf{r}_{it} \mid} \left( \mathbf{r}_{p'} - \mathbf{r}_c \right) \times \mathbf{F}_{\mathbf{r}_{pp'}},
 \end{eqnarray}

\noindent where $\mathbf{F}_{\mathbf{r}^{c,u}_b}=\pm \rho \left(\mathbf{u}_{\mathbf{r}_b^{c,u}}- \mathbf{U}_p \right)/ \triangle t$ is the force induced by covered, $\mathbf{r}^c_b$, and uncovered, $\mathbf{r}^u_b$ lattice nodes due to  due to particle motion \cite{BMS08,DL03,DA03}. Steric interaction forces, $\mathbf{F}_{\mathbf{r}_{i}}$, between the particles and between the particles and stationary solid zones, including channel walls and inline obstacles, are expressed in terms of two-body Lennard-Jones potentials \cite{BSWB18,BS10} such that $\mathbf{F}_{\mathbf{r}_{i}}=-\psi \left( \frac{\mid \mathbf{r}_i \mid} {\mid \mathbf{r_{it}} \mid} \right)^{-13} \mathbf{n}$, where $\mid \mathbf{r}_i \mid$ is the distance between a particle surface node and the neighboring particle surface node ($\mathbf{r}_{i} = \mathbf{r}_{pp'}$) or between a particle surface node and the stationary solid node on channel walls or inline obstacles ($\mathbf{r}_{i} = \mathbf{r}_{pw}$); $p$ is the particle index; $\mid \mathbf{r}_{it} \mid$ is the repulsive threshold distance; $\mathbf{n}$ is the unit vector along $\mathbf{r}_{i}$; and $\psi$ is the repulsive strength between the particles and between the particles and stationary solid nodes. 

The new position of the center of mass of a particle is computed as $\mathbf{x}_c\left(t+\triangle t \right) =\mathbf{x}_c \left( t\right)+\mathbf{U}_p\left(t \right)\triangle
t$. The population densities at $\mathbf{r}_v$ and $\mathbf{r}_v +\mathbf{e}_i \triangle t$ are updated to account for particle-fluid hydrodynamics in accordance with \cite{L94a}

\begin{equation}
 \label{e.MP7a} f^{\prime}_i\left(\mathbf{r}_v,t+\triangle t\right)=f_i(\mathbf{r}_v,t^{\ast})-\frac{2\rho \omega_i}{c_s^2}\left(\mathbf{u}_{\mathbf{r}_b} \cdot \mathbf{e}_i \right),  
 \end{equation}

\begin{equation}
 \label{e.MP7b}  f_i\left(\mathbf{r}_v+\mathbf{e}_i \triangle t,t+\triangle t\right)=f^{\prime}_i(\mathbf{r}_v+\mathbf{e}_i \triangle t,t^{\ast})+\frac{2\rho
\omega_i}{c_s^2}\left(\mathbf{u}_{\mathbf{r}_b}  \cdot \mathbf{e}_i \right).
 \end{equation}

\noindent where $f^{\prime}_i$ corresponds to population densities that propagate in $-\mathbf{e}_i$  after collision and $t^{\ast}$ represents the post-collision time. In the end of each time-step, the location of vertices on angular-shaped surfaces or boundary nodes on curved surfaces are updated using the geometrical relations in Eqs. \ref{eq_circular}-\ref{eq_rectangular}. The distance $\mathbf{d}_i=\left(d_{ix},d_{iy} \right)$ between the $i^{th}$ vertex (or a boundary node) and $\mathbf{x}_c$ is computed via $\mathbf{d_i}=\mathbf{x}_i - \mathbf{x_c} $. After $\mathbf{x}_c\left(t+\triangle t \right)$ is computed, new positions of vertices (or boundary nodes) are updated by

 \begin{equation} 
  \label{vertex_update}
 \left[ \begin{array}{c} x_i\\ y_i \end{array} \right] = \left[ \begin{array}{c} x_c \\ y_c \end{array} \right]+   \left[ \begin{array}{c} d_{ix} cos\left( \left(\Omega_p+\Upsilon_i \right) \triangle t\right) \\ d_{iy} sin \left( \left(\Omega_p+\Upsilon_i \right) \triangle t\right) \end{array} \right] 
 \end{equation}

\noindent where $\Upsilon_i$ is the angle between $\mathbf{x}_i -\mathbf{x_c}$ and $+x$. 

\section{Model Validation}\label{Validations}

\noindent \textbf{DSP-LBM}. The earlier version of the 2D LBM accommodating circular particles only successfully captured the 3D particle flow dynamics (e.g., particle velocities and trajectories) in microfluidic experiments \cite{BASD13}, by implementing Reynolds number ($Re$)-based dimensional scaling \cite{FHJ94}, given that a circular-cylindrical particle would have the same wake structure as the spherical particle, but at a lower $Re$ . The DSP-LBM was validated in \cite{BSWB18} against two benchmark problems, involving previously reported numerically-computed settling trajectories of a circular particle \cite{FHJ94}, and the trajectories and angular rotations of a settling elliptical particle in an initially quiescent Newtonian fluid \cite{ZCR09}. Ref. \cite{BSWB18} reported also successful comparisons of experimentally-determined \cite{GML71} and DSP-LBM simulated terminal velocities of a gravity-driven settling of spherical particles 5$\%$ or 10$\%$ denser than the bulk fluid. DSP-LBM simulations in \cite{BSWB18} were practically insensitive to the grid resolution. The same lattice spacing and particles' dimensions in \cite{BSWB18} were adopted in DSP-LBM simulations in this paper, in which $R_p$=10 lattice units, $N_{Bnd}$=100 for DCsPs, and the aspect ratio of the elliptical and rectangular particles is 2.  $N_{Bnd}$=100 led to a lattice spacing of $\sim0.6-0.7<1$ along the circumference of circular and elliptical particles, which eliminated the risk for missing any IPBN or EPBNs in calculating Eq. \ref{e.pfh3}.

\vspace{5mm} 

\noindent \textbf{Non-Newtonian Fluid Flow Module}.  The generalized analytic solution for the steady velocity profile of non-Newtonian ($n \neq1$) or Newtonian ($n=1$) Poiseuille flow is given by \cite{PKYPK07,W68}:

 \begin{equation}
 \label{nn6} u(y) =\left( \frac{1}{\xi} |\mathbf{g}| \right)^{\frac{1}{n}}  \left(  \frac{W}{2}\right)^{1+\frac{1}{n}} \left( \frac{n}{n+1} \right)  \left[ 1- \left(  \frac{2|y|}{W}  \right)^{1+\frac{1}{n}}  \right],
 \end{equation}

\noindent
where $W$ is the channel width and $y$ is vertical distance, orthogonal to the main flow direction, from one of the channel walls. DSP-LBM simulations of flow of a Newtonian fluid with $\nu=$1 cm$^2/$s were performed in a channel with  $W$=0.05 cm. Local kinematic viscosities for non-Newtonian fluid flow were computed from $\nu^{*}= \left[ 2^{n-1}   {| \Pi_{D} |}  ^  {\frac{n-1}{2}} \right]  \xi$. In these simulations, the maximum steady fluid velocity, $u_{max}$, was set to 9 cm/s by setting $\xi=0.85\nu$ for $n=0.8$ (pseudoplastic), $\xi=1.0\nu$ for $n=1$ (Newtonian), and $\xi=1.045\nu$ for $n=1.2$ (dilatant) fluid flows. Although $\xi$ is often determined experimentally, in DSP-LBM simulations $\xi$ for each non-Newtonian fluid was determined by calibrating the magnitude of the maximum fluid velocity against the analytic solution in Eq.\ref{nn6}, in which calibration does not affect the shape of the velocity profile, as shown in \cite{BHNS17}. DSP-LBM simulations with $n=0.8$ and $n=1.2$ were conducted to investigate the effect of small deviations from the Newtonian fluid behavior on the particle motility. The upper and lower limits for $\nu^*$ were set to $10^{-5}$ and $0.1$ to avoid unphysical values for $\nu^*$ \cite{GDK05,BHNS17}. 

\begin{figure}[h!]
\begin{center}
\scalebox{0.27} {\includegraphics{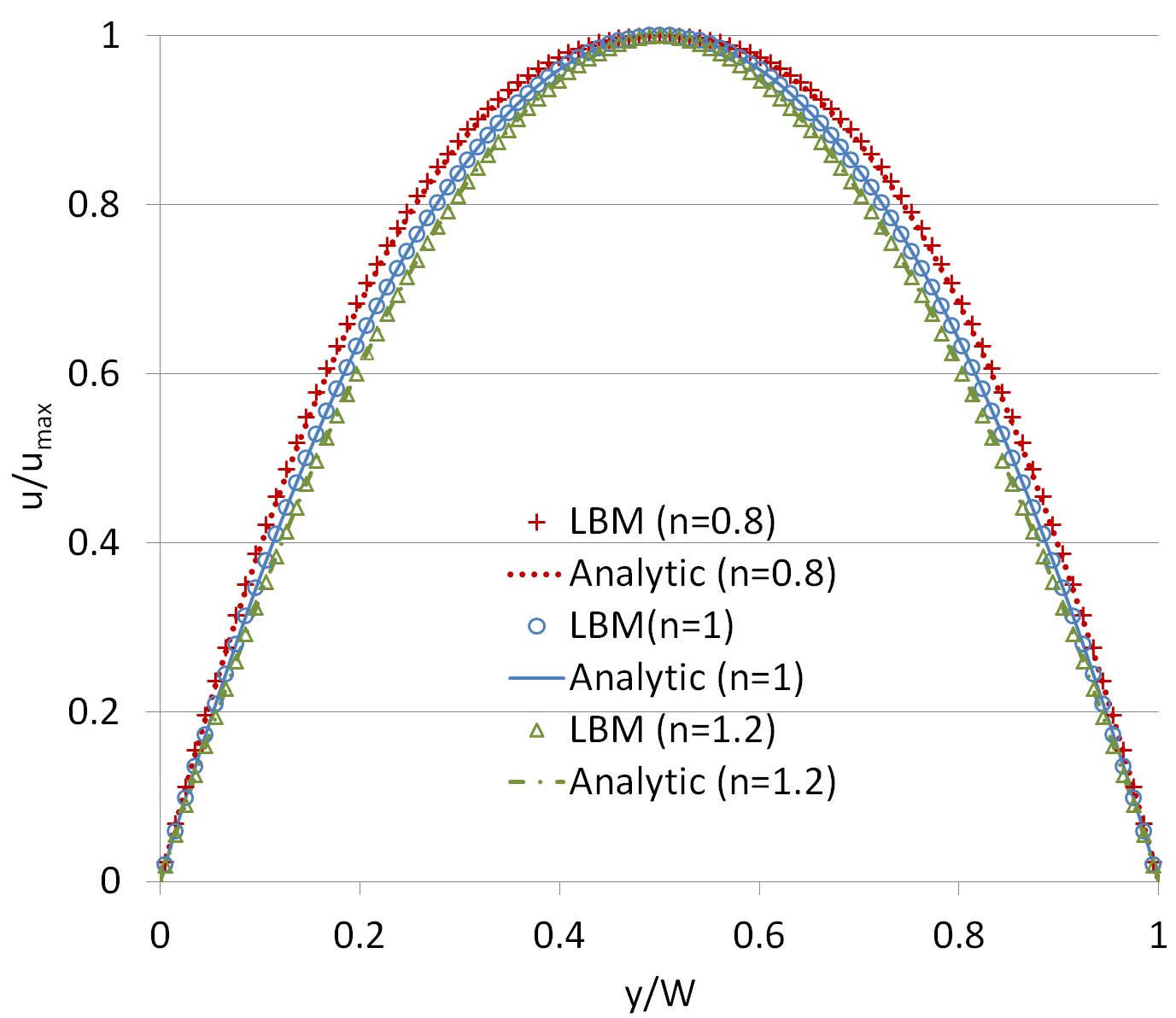}} 
\caption{Normalized steady velocity profiles of pseudoplastic $(n=0.8)$, Newtonian $(n=1)$, and dilatant $(n=1.2)$ Poiseuille flows.  Fluid velocities are normalized with respect to the maximum fluid velocity, $u_{max}$, at the midchannel. }\label{fig:Non_newtonian}
\end{center}
\end{figure}

Fig.~\ref{fig:Non_newtonian} shows that the steady velocity profiles computed by DSP-LBM closely matched the analytic solutions given by Eq. \ref{nn6}. The successful model validations, involving particle settling in a Newtonian fluid in \cite{BHNS17} and the steady non-Newtonian and Newtonian velocity profiles in Fig. \ref{fig:Non_newtonian}, suggest that the SRT is appropriate for DSP-LBM simulations; therefore, the computationally demanding multi-relaxation time was not adopted in simulations in this paper. This is consistent with the discussion in \cite{PMR16}.     

\section{DSP-LBM Simulations Involving a Single DSP}\label{1P_Simulations}

The DLD has been implemented to segregate particles based on their shapes or sizes in a Newtonian fluid in microfluidic devices with specifically designed internal geometric features \cite{BMC18,HCA04,ZRZ13}. The following research inquiries were investigated in this section: (i) how differently a circular particle would undergo lateral displacements after being released into Newtonian or non-Newtonian fluid flow in a microchannel even in the absence of geometrically complex internal geometric features? and (ii) how would particle trajectories and velocities vary with the fluid types and particle shapes? The answers to these questions would provide insights into whether the assumption of a Newtonian fluid for non-Newtonian fluids and/or the circular particle shape for non-circular particles would be reliable in optimizing microfluidic designs for entrapment or segregation of arbitrary-shaped CTCs in body fluid.        

DSP-LBM simulations with a single DSP in Newtonian or non-Newtonian Poiseuille flow were performed to address these inquiries. The channel length, $L$, (in the direction of the main flow) and width, $W$, were set to $100R_p$ and $10R_p$. A periodic boundary condition was imposed at the inlet and outlet, and a no-slip boundary condition was imposed along the channel walls. A circular, elliptical, hexagonal, or rectangular particle was released at a point 20$\%$ off the midchannel into a pseudoplastic $\left(n=0.8\right)$, Newtonian $\left(n=1.0\right)$, or dilatant $\left(n=1.2\right)$ fluid with an average steady velocity, $u_{avg}$, 6.0 cm/s prior to particle release. $R_p=500 \mu$m, $\nu=0.01$ cm$^2$/s (for a Newtonian fluid), and the resultant flow Reynolds number, $Re=u_{avg}R_p/\nu = 30$. All DSPs had the same surface area of $7.86\times10^{-3}$ cm$^2$. This simulation was set-up such that the circular particle (the reference particle) in a Newtonian fluid  drifted to an equilibrium position $\sim8\%$ off the midchannel near the outlet, exhibiting the S\'egre-Silberberg effect that a rigid spherical and neutrally buoyant particles typically displays in slow flow \cite{SS61,SS62}.

\begin{figure}[h!]
    \centering
        \begin{subfigure}{.49\textwidth}
        \centering
        \includegraphics[width=\textwidth]{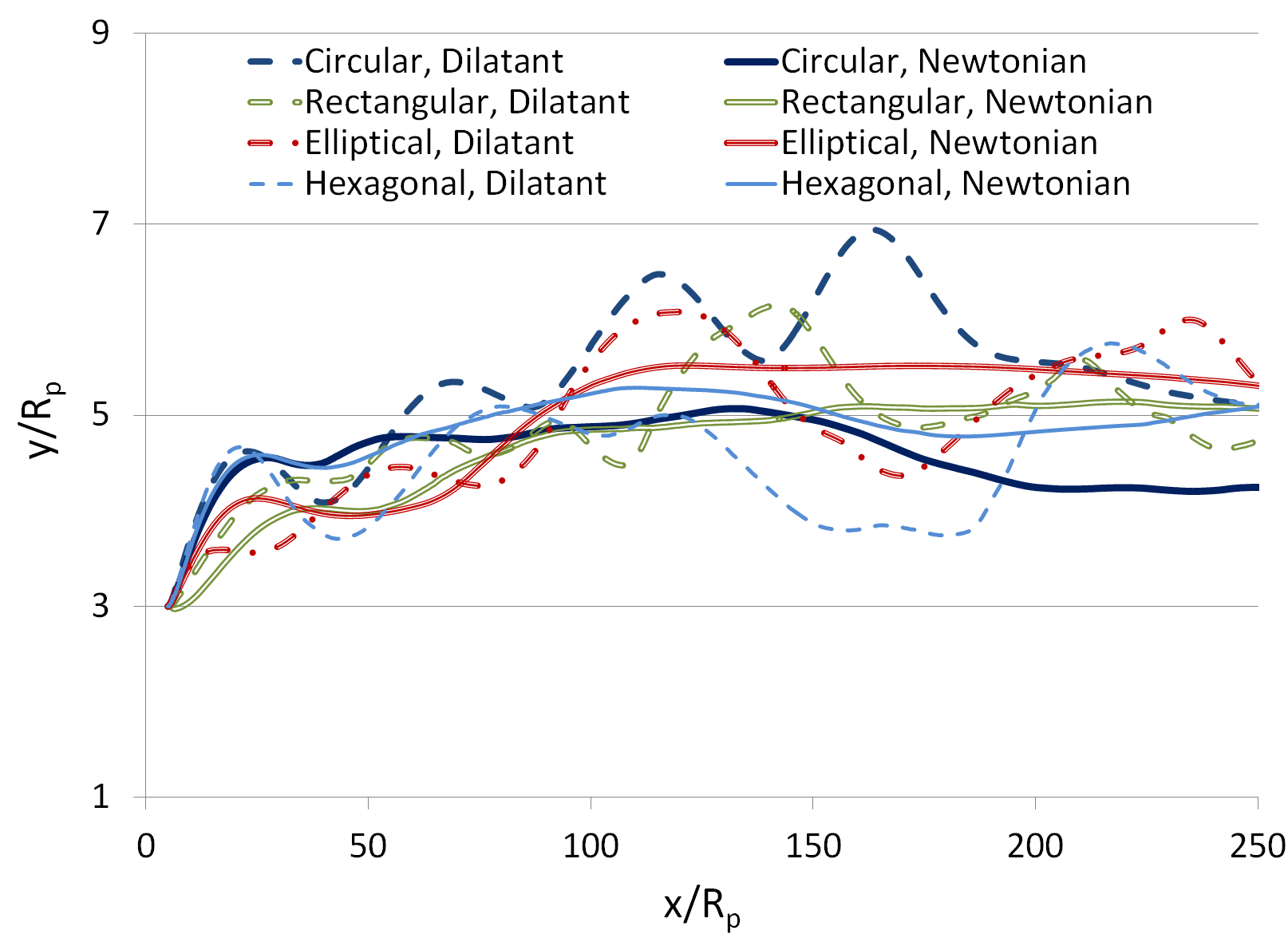}
    	\subcaption{Particle trajectories}\label{fig:DP_Tr}
    \end{subfigure}
    \hfill
    \begin{subfigure}{.49\textwidth}
        \centering
        \includegraphics[width=\textwidth]{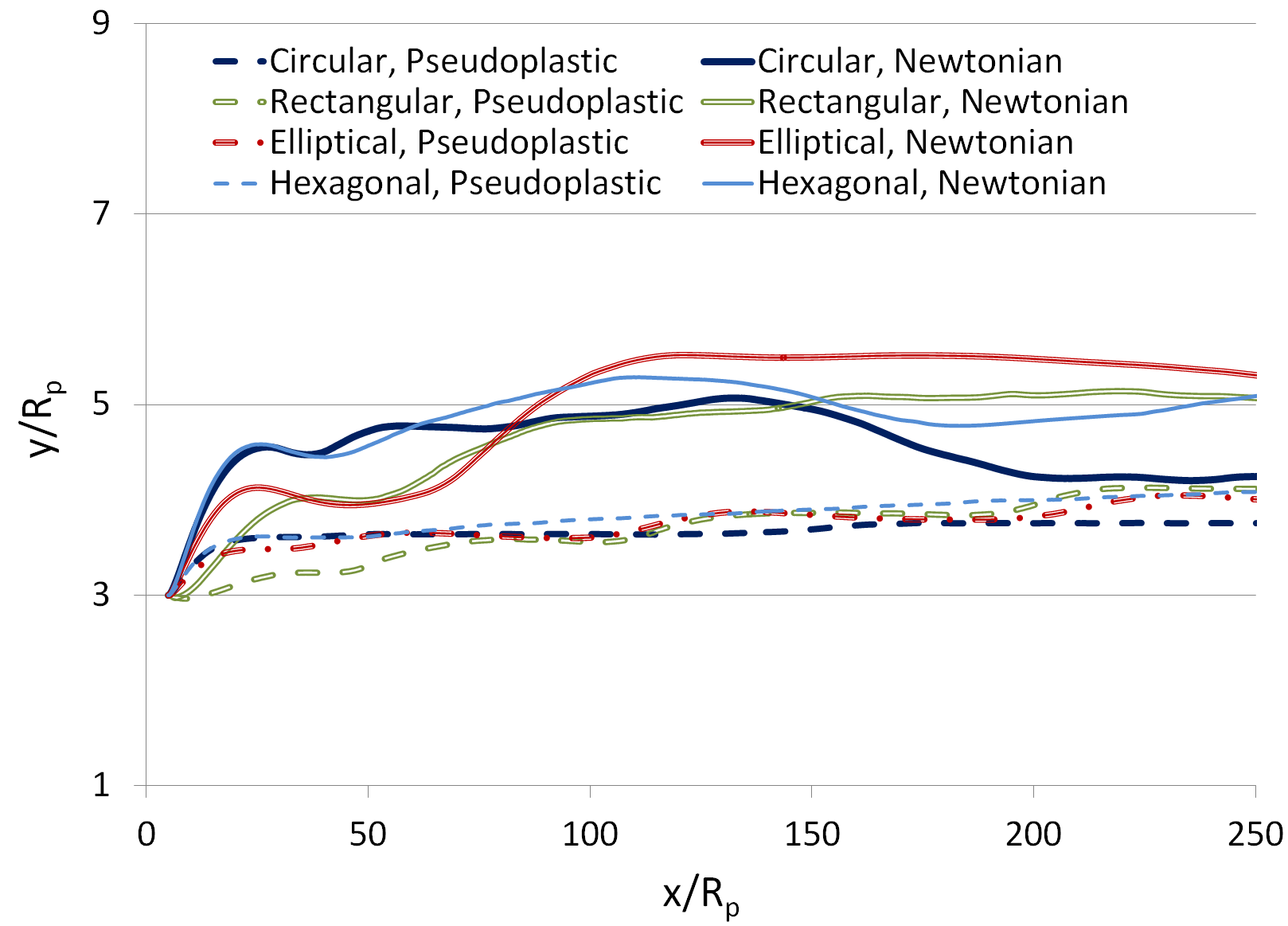}
    	\subcaption{Particle trajectories}\label{fig:DP_Vel}
    \end{subfigure}
    
    \centering
    \begin{subfigure}{.49\textwidth}
        \centering
        \includegraphics[width=\textwidth]{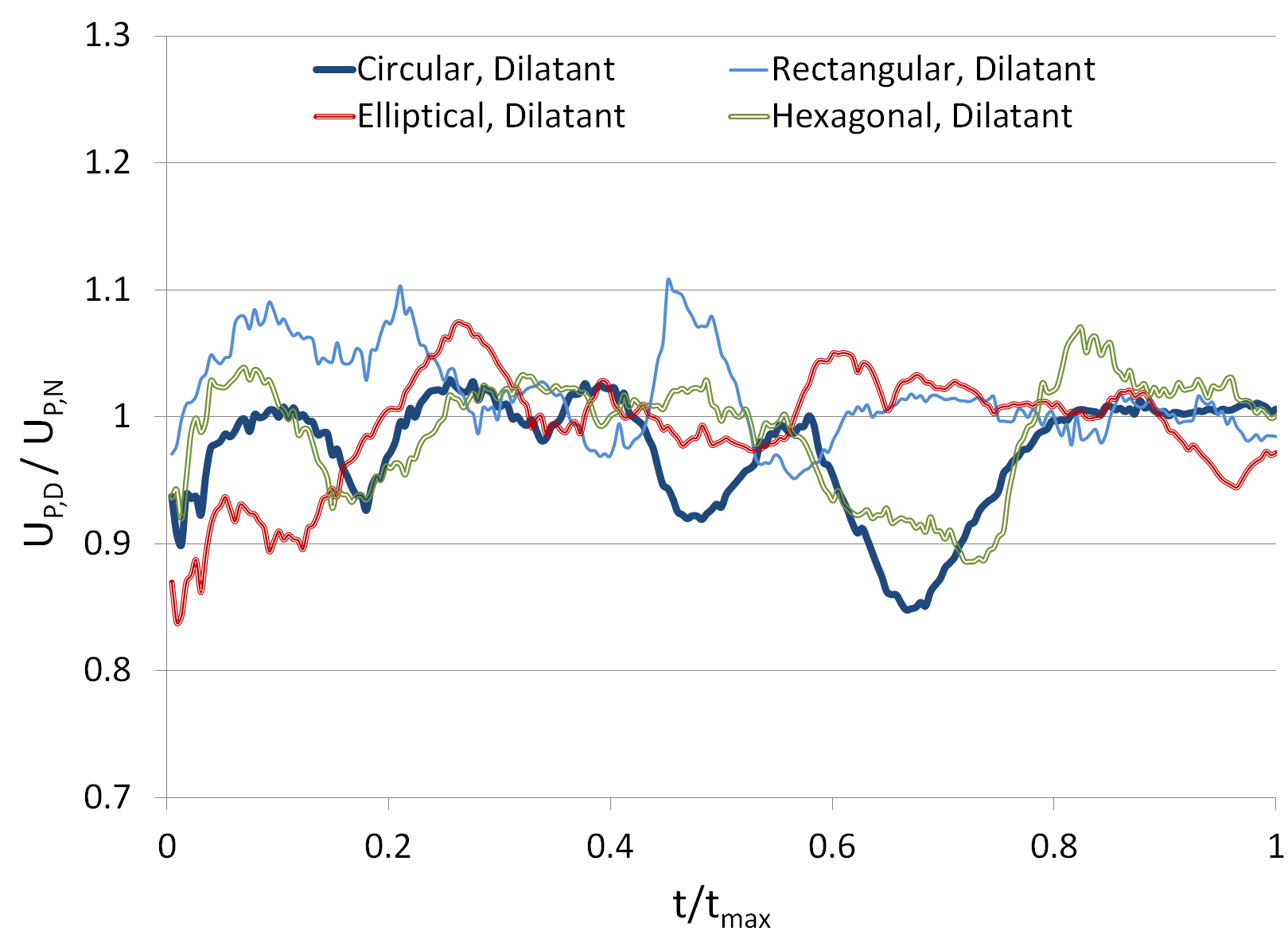}
    	\subcaption{Particle velocities}\label{fig:P_Tr}
    \end{subfigure}
    \hfill
    \centering
    \begin{subfigure}{.49\textwidth}
        \centering
        \includegraphics[width=\textwidth]{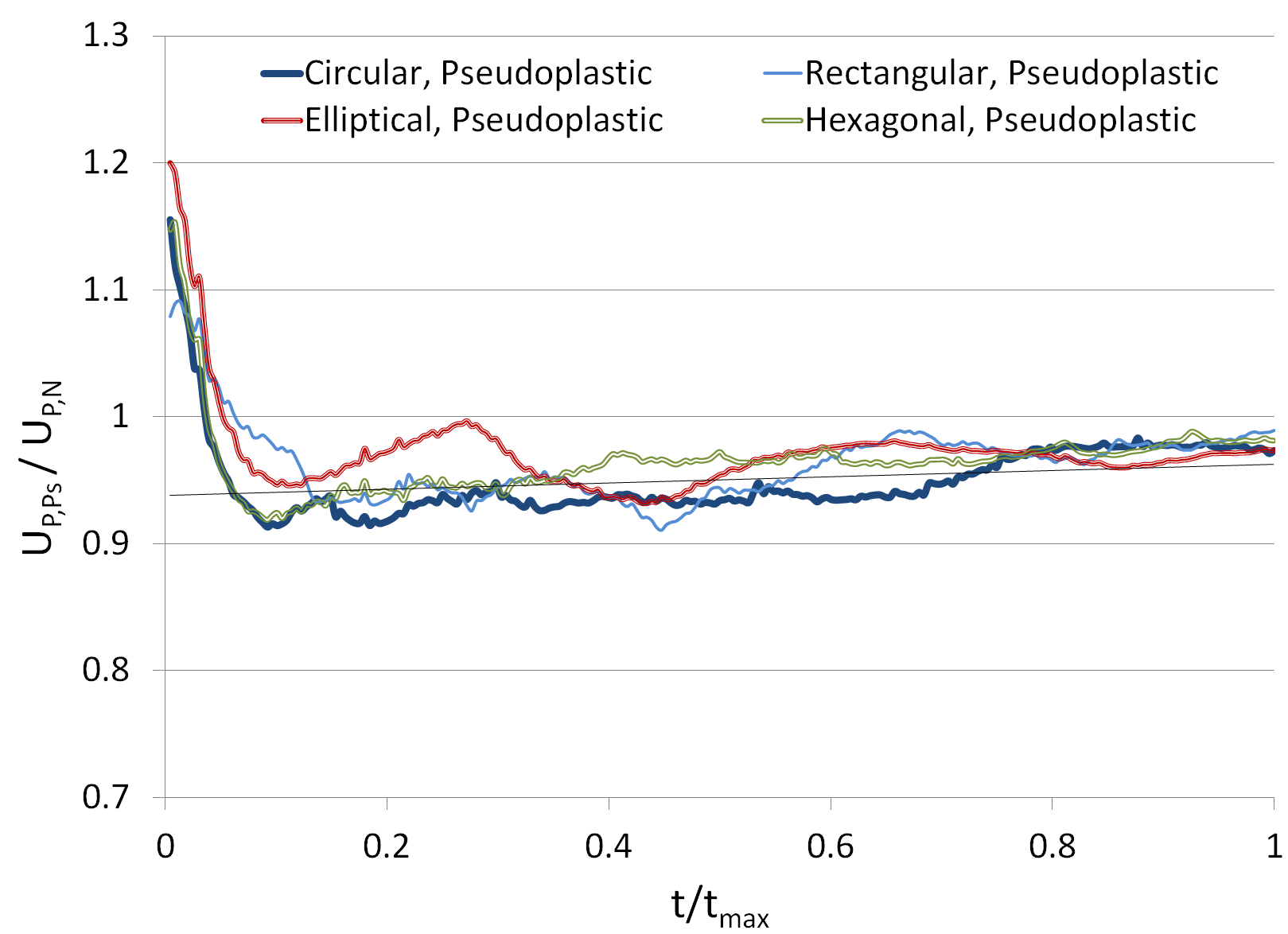}
    	\subcaption{Particle velocities}\label{fig:P_Vel}
    \end{subfigure}

    \caption{Trajectories of DSPs in (a) dilatant or Newtonian and (b) Newtonian or pseudoplastic fluid flow. The ratio of velocities of DSPs in (c) dilatant and (d) pseudoplastic fluid flow to their velocities in  Newtonian fluid flow. The first subscript of $U$ denotes the particle shape and the second subscript denotes the fluid type (D, N, and Ps corresponds to dilatant, Newtonian, and pseudoplastic fluids). $U_{CD}$, for example, represents the velocity of a circular particle in a dilatant fluid.    \label{fig:Single_particle}}
\end{figure}

In Figs. \ref{fig:DP_Tr}-d, DSPs displayed steady equilibrium with monotonic drift to the equilibrium position away from the midchannel in pseudoplastic fluid flow, steady equilibrium with transient overshoot in Newtonian fluid flow, and weak to strong oscillatory motion in dilatant fluid flow in a 2D microchannel when $u_{avg}$ was the same in all simulations. The circular particle exhibited larger oscillations and transient overshoot due in part to its smaller $I_p$ than the other particles, which led to the least resistance to angular rotations. In contrast, the rectangular and elliptical particles have higher $I_p$'s, and hence, they exhibit relatively stronger resistance to rotations. Although the hexagonal particle has slightly higher $I_p$ than the circular particle, it exhibits persistent rotations as it travels along its equilibrium position in a channel \cite{BSWB18} due to asymmetric position of its vertices about its equilibrium position.

Among these simulations, the DSP experienced the highest inertial effects in dilatant fluid flow due to the largest fluid velocity differentials on its opposite surfaces orthogonal to the main flow direction as it approached and migrated around the midchannel (Fig. \ref{fig:Non_newtonian}). Such transitions in migration modes of DSPs are similar to changes in trajectories of gravity-driven settling of a circular particle when the inertial effects were elevated by increasing the particle density \cite{FHJ94,BSWB18} or changes in trajectories of neutrally buoyant circular or DSPs flowing in a Newtonian fluid with higher flow strengths \cite{BMS08,BASD13}. Thus, elevated inertial effects on the trajectories of a mobile particle in a dilatant fluid caused by relatively sharper gradients of fluid velocities than in Newtonian or pseudoplatic fluids are comparable to the elevated inertial effects on the settling trajectories of a denser particle in an initially quiescent fluid or flow trajectories of a particle in higher $Re$-flow. Overall transitions in flow trajectories of a single DSP in pseudoplastic fluid flow to flow trajectories in dilatant fluid flow are consistent with settling trajectories of a circular particle \cite{FHJ94}, or an elliptical particle \cite{ZCR09}, or a particle of other geometric shapes as the particle density is increased \cite{BSWB18}. 

In Figs. \ref{fig:DP_Tr}-d, after arriving at $x/R_p \sim 60$, the circular particle traveled in the left side of the channel $\left( y/R_p>5 \right)$ in a dilatant fluid; whereas, it remained in the right side of the channel in Newtonian or pseudoplastic fluids. If a dilatant fluid was approximated in numerical simulations by or replaced in microfluidic experiments with a Newtonian fluid, the lateral displacement of the circular particle would be underestimated, for example, by a distance of $2.1R_p$ (=0.21$W$) at $x/R_p \sim 160$. Similarly, the lateral displacement of the circular particle would be overestimated by a distance of $1.1R_p$ (=0.11$W$) at $x/R_p \sim 160$, if a pseudoplastic fluid was approximated by a Newtonian fluid. Strong sinusoidal oscillations in the trajectory of the circular particle in a dilatant fluid at early times was damped in Newtonian and pseudoplastic fluids. Although all DSPs were drifted to the vicinity of the midchannel ($y/R_p=5$ at $x/R_p=250$) in a dilatant fluid, only non-circular particles were drifted to the midchannel in a Newtonian fluid at $x/R_p=250$. In pseudoplastic and Newtonian fluid flows, the circular and non-circular particles drifted to different quasi-steady equilibrium positions at $x/R_p=250$, revealing the sensitivity of the equilibrium position of a particle to its geometric shape. 

Figs. \ref{fig:DP_Tr}-d also revealed that the S\'egre-Silberberg effect is not only related to the particle size-based $Re$, but also to the fluid type and particle shape. Unlike in dilatant fluid flow, all DSPs exhibited the S\'egre-Silberberg effect in pseudoplastic fluid flow. However, only the circular particle displayed the S\'egre-Silberberg effect in the Newtonian fluid flow.

The ratio of the velocity of a particular DSP in a dilatant fluid to its velocity in a Newtonian fluid in Figs. \ref{fig:P_Tr}-d shows that approximating a dilatant fluid with a Newtonian fluid would result in errors up to of 3-11$\%$ (the largest for the elliptical particle and the smallest for the rectangular particle) in local particle velocities. Similarly, if pseudoplastic fluid flow is approximated by non-Newtonian fluid flow, errors in local particle velocities would be up to 9-20$\%$ (the largest for the rectangular particle and the smallest for the circular particle). The travel times of the circular, hexagonal, and elliptical particles were 2.8$\%$, 1.1$\%$, and 0.8$\%$ longer in dilatant fluid flow than in Newtonian fluid flow. In contrast, the rectangular particle traveled 1.5$\%$ faster in a Newtonian fluid than in a dilatant fluid. On the other hand, the travel times of DSPs were 2.9-5.3$\%$ longer (being the longest for the circular particle and the shortest for the elliptical particle) in pseudoplastic fluid flow than in Newtonian fluid flow.  

In brief, DSP-simulations with a single DSP show that that particle trajectories, lateral displacements, velocities, and travel times vary with the fluid type and geometric shape of the particles. These simulations, even without considering the effect of interparticle interactions and inline obstacles, demonstrated that inaccurate representation of the actual geometry of the particles and non-Newtonian behavior of the fluid could result in 1.1-2.1$R_p$ errors in lateral displacements of particles, up to 3-20$\%$ errors in particle velocities, and 3-5$\%$ errors in travel times of DSPs.

\section{DSP-LBM Simulations with a Mixture of DSPs in a Microchannel}\label{4P_Simulations}

\begin{figure}[h!]
\begin{center}
\scalebox{0.3} {\includegraphics{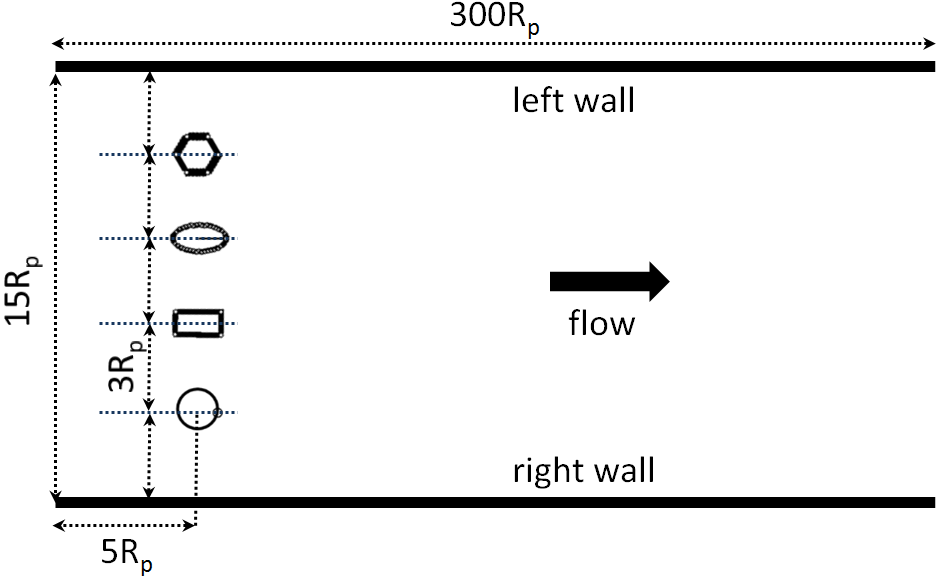}} 
\caption{A schematic representation of the microfluidic channel geometry and release locations of particles in DSP-LBM simulations. All particles have the same surface area of $7.86\times10^{-3}$ cm$^2$. The initial tilt angle of non-circular particles is 0$^\circ$. $\psi$=1 and $\mathbf{r_{it}}$=2.5 lattice spacing in simulating particle-particle and particle-wall interactions \cite{BSWB18}. }\label{fig:flowDomain_4P}
\end{center}
\end{figure}

DSP-LBM simulations were performed to investigate the combined effects of the fluid type and flow strength on the flow trajectories and velocities of a mixture of DSPs in a microchannel (Fig. \ref{fig:flowDomain_4P}) by accommodating interparticle interactions in Eq. \ref{e.6}. Because the blood is considered as a pseudoplastic fluid, only a pseudoplastic fluid is used for non-Newtonian fluid flow simulations in the following sections.

The properties of the fluid and particles, and the specifications of the flow domain boundaries in simulations in Section \ref{1P_Simulations} were adopted for simulations in this section,  except for the channel width. The channel width was widened from $10R_p$ to $15R_p$ to initially place 4 particles with the center-to-center separation distance of $3R_p$ (Fig. \ref{fig:flowDomain_4P}). The initial separation distance of $3R_p$ was chosen to be slightly longer than the length of the long axis of the elliptical particle ($2.8R_p$) and the long side of the rectangular particle ($2.5R_p$). Different from simulations in Section \ref{1P_Simulations}, circular, rectangular, elliptical, and hexagonal particles (CREH configuration) were simultaneously released from a multiple-port near the inlet into the fluid after the steady flow field was established. DSP-LBM simulations were  performed with a pseudoplastic or Newtonian fluid flowing in a microchannel with an average steady velocities of 6.0 cm/s, corresponding to $Re=30$, (slow flow) and 12.0 cm/s, corresponding to $Re=60$, (fast flow) prior to releases of DSPs into the fluid. This simulations were set up such that the circular particle (the reference particle) drifted monotonically to the midchannel at low inertial effects at $Re=30$, but its trajectory displayed transient overshoot about the midchannel at higher inertial effects at $Re=60$ \cite{BMS08,FHJ94,BASD13}.

\begin{figure}[h!]
    \centering
    \begin{subfigure}{.49\textwidth}
        \centering
        \includegraphics[width=\textwidth]{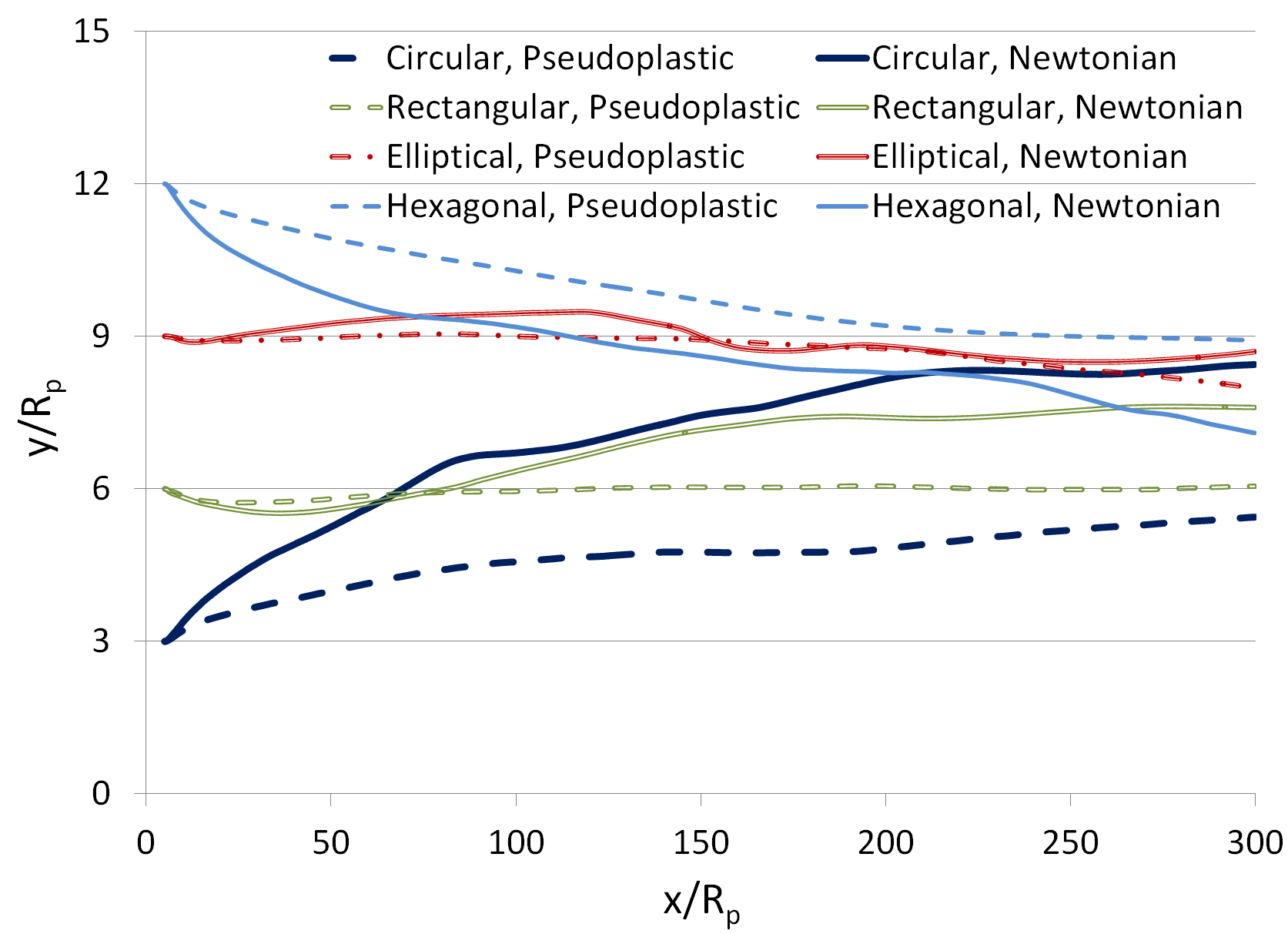}
    	\subcaption{Particle trajectories in a slow flow}\label{4P_Slow_Tr}
    \end{subfigure}
    \hfill
    \begin{subfigure}{.49\textwidth}
        \centering
        \includegraphics[width=\textwidth]{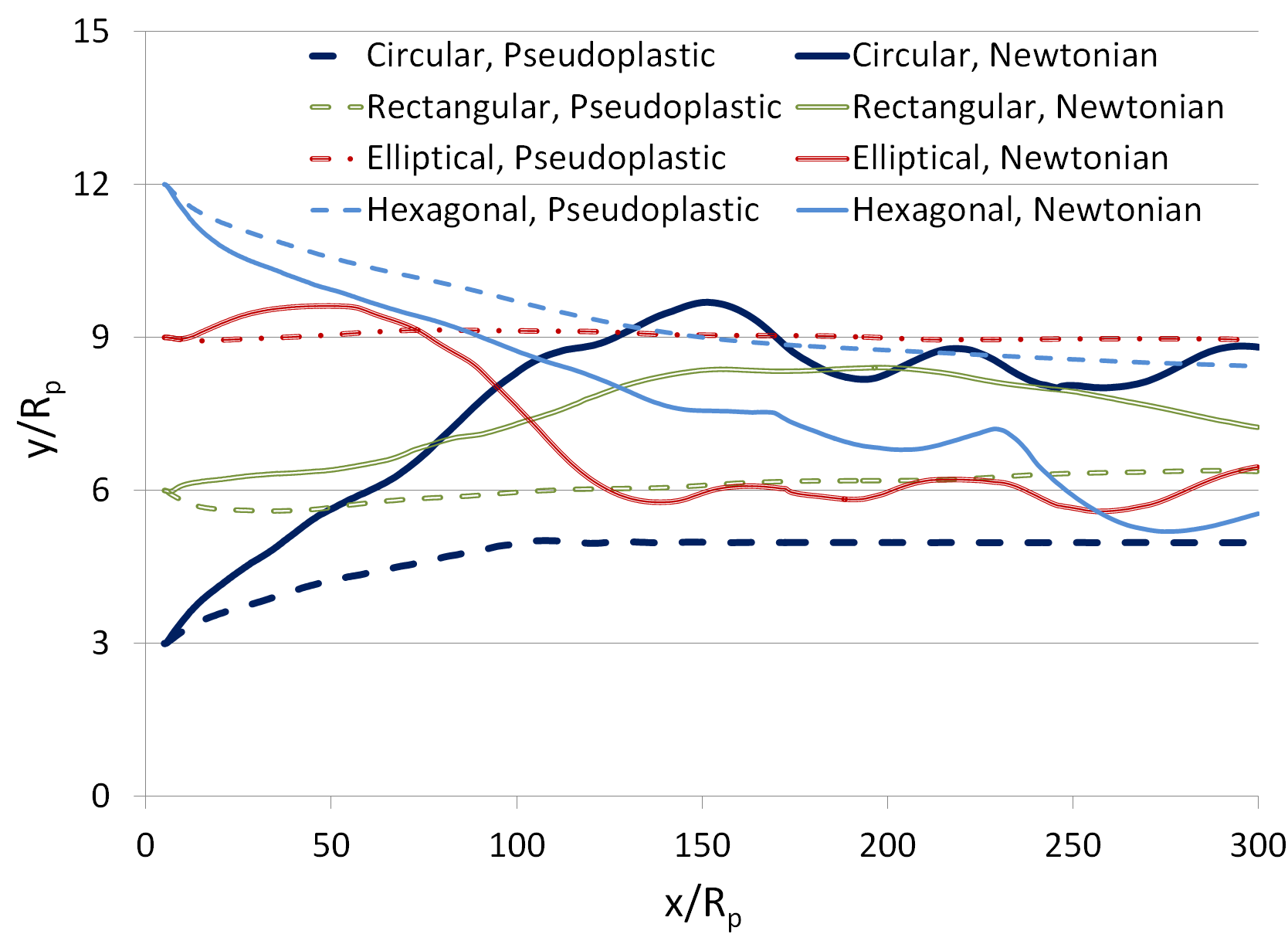}
    	\subcaption{Particle trajectories in a fast flow}\label{4P_Fast_Tr}
    \end{subfigure}
    
    \centering
    \begin{subfigure}{.49\textwidth}
        \centering
        \includegraphics[width=\textwidth]{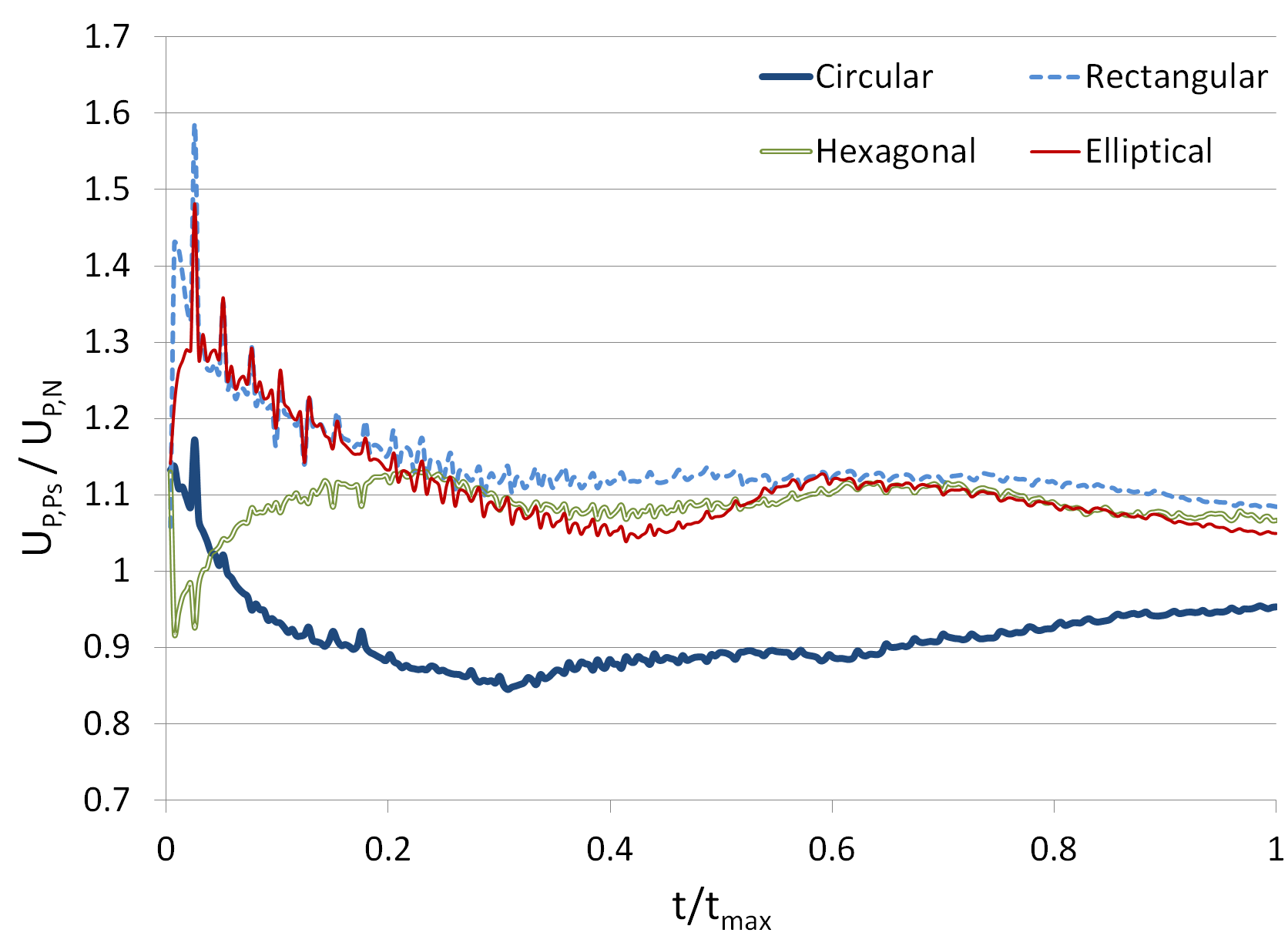}
    	\subcaption{Particle velocities in a slow flow}\label{4P_Slow_Vel}
    \end{subfigure}
    \hfill
    \centering
    \begin{subfigure}{.49\textwidth}
        \centering
        \includegraphics[width=\textwidth]{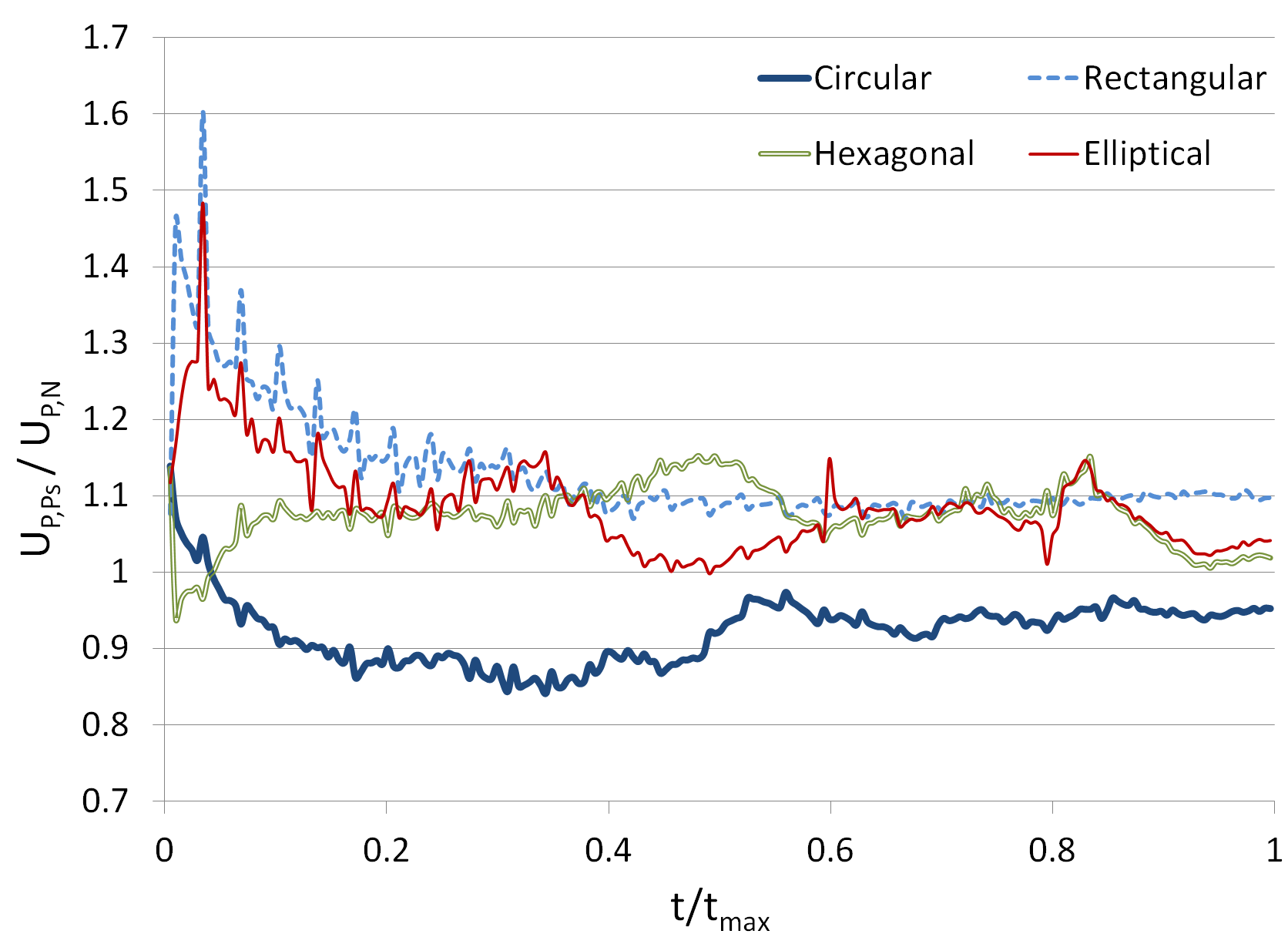}
    	\subcaption{Particle velocities in a fast flow}\label{4P_Fast_Vel}
    \end{subfigure}
       
 \caption{Flow trajectories of DSPs in pseudoplastic and Newtonian fluids with an average steady velocity of (a) slow flow ($Re=30$) and (b) fast flow ($Re=60$). The ratio of velocities of DSPs in a pseudoplastic fluid to their velocities in a non-Newtonian fluid in (c) slow flow ($Re=30$) and (d) fast flow ($Re=60$). \label{fig:4_DSP}}
\end{figure}

At $Re=30$, DSPs drifted toward the midchannel in Fig. \ref{4P_Slow_Tr} due to the combined inertial and wall effects. Particles' drifts were more pronounced in a Newtonian fluid flow as the velocity gradients of the fluid around the midchannel were sharper in a Newtonian fluid than in a pseudoplastic fluid (Fig. \ref{fig:Non_newtonian}). As a result, DSPs drifted to semi-equilibrium positions off the midchannel in a pseudoplastic fluid. However, at $Re=60$ with higher inertial effects, DSPs displayed oscillatory trajectories with transient overshoot about the midchannel (Fig. \ref{4P_Fast_Tr}) in a Newtonian fluid, in an agreement with the behavior of circular particle in a Newtonian fluid with high inertial effects in \cite{FHJ94}. On the contrary, the flow trajectories of DSPs in a pseudoplastic fluid were less sensitive to the flow strengths considered here. Relatively low velocity gradients orthogonal to the main flow direction in a pseudoplastic fluid was not sufficient to impose large uneven hydrodynamic forces on the opposite sides of the particle for the particle to gain sufficient angular momentum to drift to the midchannel. Thus, when DSPs were away from walls, they displayed small lateral displacements in a pseudoplastic fluid, regardless of the flow strengths considered.             

Additional observations from Figs. \ref{4P_Slow_Tr}-d are noteworthy. At $Re=30$, the trajectories of the elliptical particle in pseudoplastic and Newtonian fluids were similar with the resultant spatial discrepancies in its lateral displacements within $R_p$. Conversely, the trajectories of the remaining particles in pseudoplastic and Newtonian fluids exhibited larger disparities as high as 1.6$R_p$ for the rectangle, 1.8$R_p$ for the hexagonal, and 3.3$R_p$ for the circular particles. Therefore, if pseudoplastic fluid flow is approximated by Newtonian fluid flow in these simulations, the maximum error in lateral displacements of the particles would be in the range of 0.1$W$ to 0.2$W$. Although the trajectories of the elliptical particle in pseudoplastic and Newtonian fluids were practically identical at $Re=30$, its lateral displacements differed as high as 3.4$R_p$ at $Re=60$ due to larger angular momentum the elliptical particle exhibited at $Re=60$. In contrary to the elliptical particle, the trajectories of the rectangular particle in pseudoplastic and Newtonian fluids at $Re=30$ and at $Re=60$ were similar. Because the elliptical and rectangular particles have the same aspect ratio and their initial release locations were symmetric about the midchannel, the particles' shape and their release positions (Fig. \ref{fig:flowDomain_4P}) were critical for their subsequent migration pathways. Fig. \ref{fig:4_DSP}c-d shows also that shortly after the particles were released, only the circular particle traveled slower in a pseudoplastic fluid than in a Newtonian fluid at both $Re=30$ and $Re=60$.  

\section{DSP-LBM Simulations with a Mixture of DSPs in a Microchannel with Inline Obstacles}\label{8P_Simulations}

\begin{figure}[h!]
    \begin{subfigure}{.75\textwidth}
        \centering
        \includegraphics[width=\textwidth]{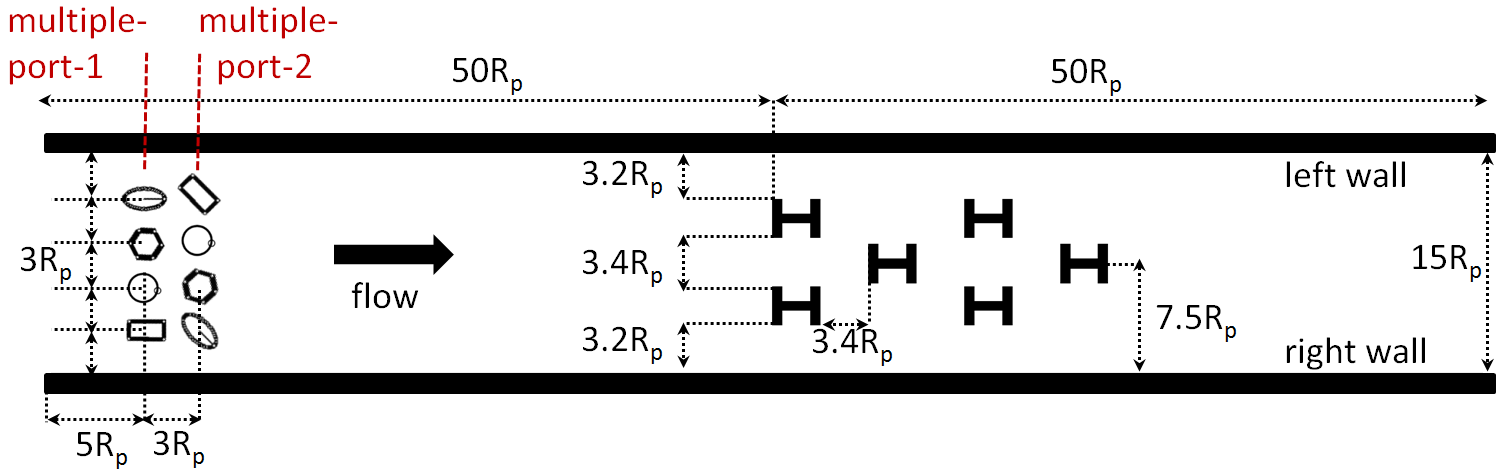}
    	\subcaption{}\label{Domain_Geometry_2}
    \end{subfigure}
    \hfill
    \begin{subfigure}{.24\textwidth}
        \centering
        \includegraphics[width=\textwidth]{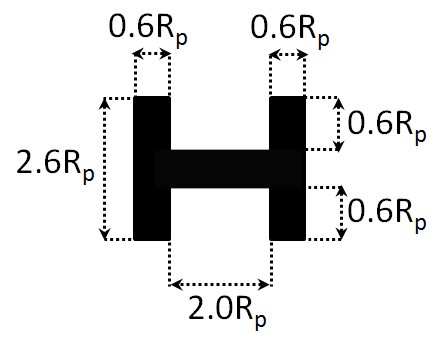}
    	\subcaption{}\label{obstacle}
    \end{subfigure}

 \caption{(a) A schematic representation of the inertial microfluidics geometry and the release locations of particles in DSP-LBM simulations, (b) geometric specifics of an I-shaped inline obstacle. $R_p$ is the radius of the circular particle.  All particles have the same surface area of $7.86\times10^{-3}$ cm$^2$. \label{fig:8_DSP}}
\end{figure}

DSP-LBM simulations were performed to investigate the combined effects of the fluid type, inline obstacles, and the order of particles' position at the release location on the flow trajectories of a mixture of DSPs in microfluidics with I-shaped inline obstacles. The geometric peculiarities of the microfluidics and inline obstacles are shown shown in Fig. \ref{fig:8_DSP}. The properties of the particles and fluids, flow domain boundaries, and the slow flow field condition in simulations in Section \ref{4P_Simulations} were adopted for DSP-LBM simulations in this section. An array of I-shaped inline obstacles was originally used in \cite{ZRZ13} for shape-based segregation of particles through the DLD method. Horizontal and vertical separation distances between inline obstacles in Fig. \ref{fig:8_DSP} were chosen to be larger than the long axis of the elliptical particle ($2.8R_p$) and the long side of the rectangular particle ($2.5R_p$) to avoid particle filtration. The particles in the second multiple port were initially rotated by 45$^\circ$ to reflect potential uncertainties associated with the initial orientations of DSPs. 

After the steady flow field was established, DSPs were released simultaneously into the fluid near the inlet from two multiple-ports (Fig. \ref{fig:8_DSP}). We considered two scenarios to investigate the effects of the order of DSPs at the release location on the particles' trajectories in pseudoplastic and Newtonian fluid flows. In the first scenario, the first multiple-port closer to the inlet involved rectangular, circular, hexagonal, and elliptical particles (the RCHE configuration) with the center-to-center separation distance of $3R_p$ from the right wall to the left wall. In the second multiple-port, the order of release positions of DSPs was flipped and all DSPs were tilted by 45$^\circ$ in the clockwise direction. This scenario will be referred to as `RCHE' hereafter, in reference to the particles arrangement in the first multiple-port.  In the second scenario, the first multiple-port involved circular, rectangular, elliptical, and hexagonal (the CREH configuration) particles with the center-to-center separation distance of $3R_p$ from the right wall to the left wall. In the second multiple port, the order of release positions of DSPs were flipped and all DSPs were tilted by 45$^\circ$ in the clockwise direction. This scenario will be referred to as `CREH' hereafter. DSPs in the following discussion are labelled with the first letter of their geometric shape followed by a number indicating from which multiple-port they are released.

\begin{figure}[ht]
    \centering
    \begin{subfigure}{.95\textwidth}
        \centering
        \includegraphics[width=\textwidth]{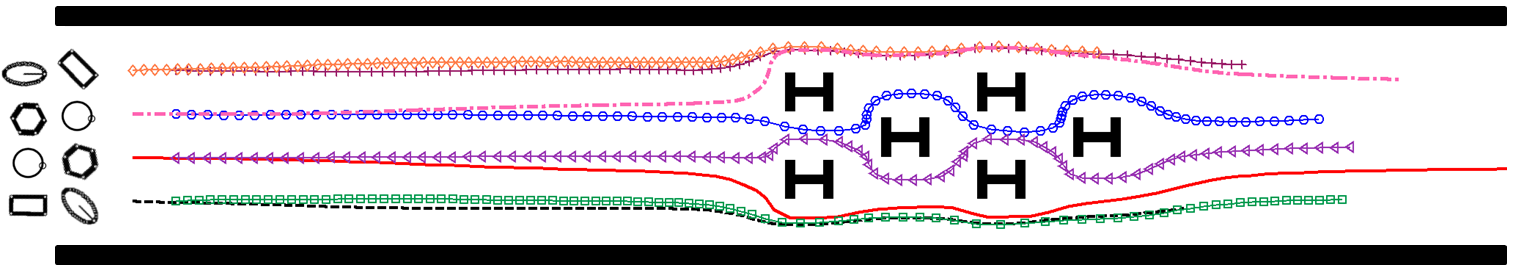}
    	\subcaption{RCHE in pseudoplastic fluid flow}\label{CREH_8P_Pseudo}
     \end{subfigure}
    \vfill
    \begin{subfigure}{.95\textwidth}
        \centering
        \includegraphics[width=\textwidth]{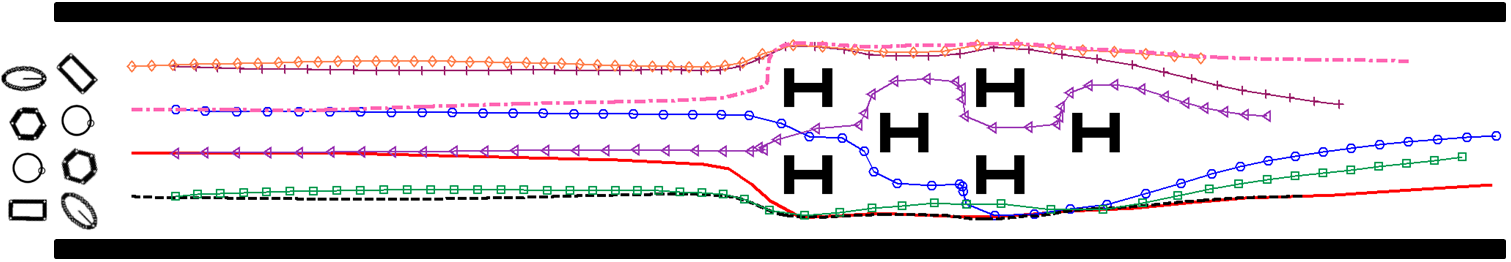}
    	\subcaption{RCHE in Newtonian fluid flow}\label{CREH_8P_Newt}
    \end{subfigure}
    \vfill
    \centering
    \begin{subfigure}{.95\textwidth}
        \centering
        \includegraphics[width=\textwidth]{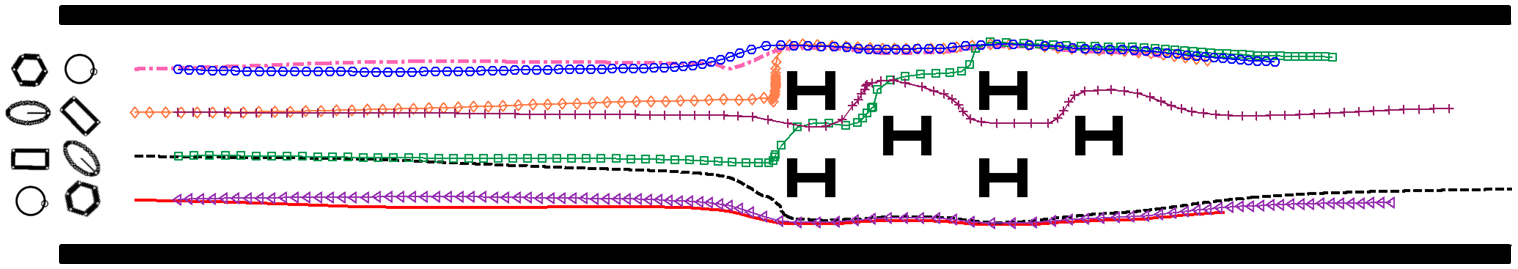}
    	\subcaption{CREH in pseudoplastic fluid flow}\label{RCHE_8P_Pseudo}
    \end{subfigure}
    \hfill
    \centering
    \begin{subfigure}{.95\textwidth}
        \centering
        \includegraphics[width=\textwidth]{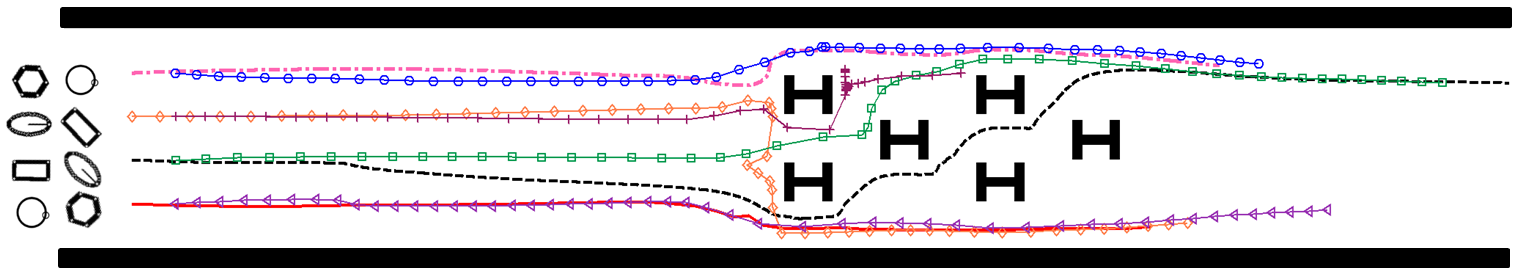}
    	\subcaption{CREH in Newtonian fluid flow}\label{RCHE_8P_Newt}
    \end{subfigure}

    \begin{subfigure}{.55\textwidth}
        \centering
        \includegraphics[width=\textwidth]{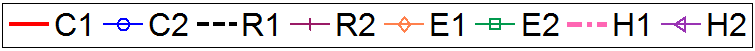}
        \label{legend}
    \end{subfigure}

    \caption{Flow trajectories of DSPs in pseudoplastic or Newtonian fluids in inertial microfluidics with inline obstacles until one of the DSPs reached the exit-end. DSPs in two different configurations (RCHE or CREH) were released after the steady flow field was established. The average steady velocity was the same in pseudoplastic and Newtonian fluid flows, corresponding to the slow flow field \textcolor{red}{$Re=30$} in Fig. \ref{fig:4_DSP}.      \label{fig:Eight_particle}}
\end{figure}

DSP-LBM simulations in Fig. \ref{fig:Eight_particle} demonstrate geometric shape-based separation of particles in both pseudoplastic and Newtonian fluids. In both the CREH and RCHE configurations, most of the particles segregated toward the left wall were angular-shaped; whereas, most of the particles segregated toward the right wall were curve-shaped. However, in pseudoplastic fluid flow, geometric shape-based particle segregation was sensitive to the initial configuration of the particles. For example, most of the particles segregated toward the left wall were angular-shaped for the RCHE configuration; whereas, the opposite was true for the CREH configuration.

Fig. \ref{fig:Eight_particle} also show that the particles released from the ports closer to the walls kept moving closer to the walls and avoided flow pathways in between inline obstacles, regardless of the fluid type and geometric shape of the particles. Conversely, the particles released from the ports closer to the midchannel exhibited different migration patterns in Newtonian and pseudoplastic fluids, which varied also with the geometric shape of the particles. The RCHE configuration in a pseudoplastic fluid (Fig. \ref{fig:Eight_particle}a) resulted in most symmetric particles' trajectories, in which three of the particles (E1, R2, C2) were separated toward the left wall and the other three (E2, R1, C1) were separated toward the right wall as they passed through the zone of obstacles while the remaining two particles (C2,H2) exhibited nearly symmetric flow trajectories between the inline obstacles.

Unlike in the pseudoplastic fluid flow, the flow trajectories of C2 and H2 in a Newtonian fluid crossed over near the first two obstacles as C2 and H2 separated to the opposite half of the channel with respect to their release locations. Subsequently, C2 gradually separated toward the right wall as it passed through the zone of obstacles while H2 migrated in between the obstacles. If pseudoplastic fluid flow was approximated by Newtonian fluid flow, the lateral displacements of C2 and H2 would be off by 6.4$R_p$ (=0.43$W$) and 5.1$R_p$ (=0.34$W$) at $x=75R_p$, outside the zone of obstacles. Such deviations would be non-negligible errors in lateral displacements of the particles in inertial microfluidics. Moreover, across the zone of the obstacles, more particles were separated toward the right wall in a Newtonian fluid than in a pseudoplastic fluid. The circular particles (C1 in the pseudoplastic fluid flow, and C1 and C2 in Newtonian fluid flow) were the fastest moving particles. The travel time of the particles in the pseudoplastic and Newtonian fluid flow differed by a factor of 0.91-1.11 for the RCHE and 0.77-1.15 for the CREH configurations.     

The effects of the order of the particles at the release location on the particles' trajectories in a pseudoplastic fluid is evident from Fig. \ref{fig:Eight_particle}a-c. Unlike C2 and H2 in the RCHE configuration, R2 and E2 released from the second multiple-port near the midchannel in the CREH configuration did not display symmetric flow trajectories. Nor did the particles evenly separated to the walls in the RCHE configuration. Two rectangular particles, R1 and R2, were the fastest moving particles in a pseudoplastic fluid flow in the CREH configuration that had the same release positions with the fastest moving C1 and C2 in the RCHE configuration. Thus, the release position of the particles appear to be more critical than their shape in determining the maximum travel time of the particles in pseudoplastic fluid flow in these simulations.

The order of the release positions of the particle was also important in Newtonian fluid flow (\ref{fig:Eight_particle}b-d), where more particles were segregated toward the left wall with negligible subsequent drifts towards the midchannel for the CREH configuration than for the RCEH configuration. Although the assumption of Newtonian fluid for the pseudoplastic fluid would lead to small error of $\sim 1 R_p$ (0.1W) in the lateral displacement of E2 at $x=75R_p$, the error in the lateral displacements of the R2, R1, and E1 outside the zone of obstacles would be as high as 4.1$R_p$ (=0.27W), 9.9$R_p$ (=0.66W), and 11.6$R_p$ (=0.78W). Hence, both the geometric shape and the fluid type could have significant effects on the flow trajectories, lateral displacements, and travel times of DSPs in inertial microfluidics with inline obstacles. Therefore, negligence of the pseudoplastic nature of the fluid and/or the exact geometric shape of the cells (or surrogate particles) would not be practical in assessing or optimizing geometric design of microfluidics proposed or designed for cell segregation. The DSP-LBM with non-Newtonian fluid flow simulation capabilities circumvents such problems in practice.        

\section{Summary and Conclusions}\label{sec:summary}

Recent numerical analyses \\cite{PCD17,HC17,KMFD16,MSAM12,JSH15,DTF15,SAM18} to assess the performance of particular microfluidic device designs in separating CTCs from healthy cells have been reported without  concurrently accommodating non-circular geometric shapes of CTCs and the non-Newtonian behavior of body fluids. We upgraded the DSP-LBM \cite{BSWB18} for non-Newtonian fluid flow simulations and used it to numerically investigate the reliability of the assumptions of a Newtonian fluid for pseudoplastic fluids and circular-shape for non-circular particles in assessing the performance of microfluidic devices with simplified geometries for shape-based segregation of particles.    

Numerical results demonstrated that the S\'egre-Silberberg that the neutrally buoyant particles were previously shown to display in slow flow is not only associated with the flow strength ($Re$), but also with the combination of particle shape and fluid type. The simulations demonstrated that if a smooth-walled channel filled with a dilatant, Newtonian, or pseudoplastic fluid flowing with the same average fluid velocity, the DSP would experience higher inertial effects in a dilatant fluid, as opposed to lower inertial effects in a pseudoplastic fluid.  

The results also revealed that the lateral displacements, velocities, and travel times of individual particles differed due to the geometric shape of the particle and the fluid type (Newtonian vs. non-Newtonian). The aforementioned assumptions resulted in errors as high as 0.21$W$ in lateral displacements of the particles in a microchannel. Simulations with a mixture of different-shaped particles in a microchannel showed that the lateral displacements of some of the particles in a mixture were practically insensitive to the fluid type. Moreover, unlike a mixture of particles in a Newtonian fluid, the lateral displacements of particles in a pseudoplastic fluid were nearly insensitive to an increased inertial effect. Yet, when these assumptions were applied, errors in the lateral displacements of the particles varied in the range of 10-20$\%$ of the channel width. Although the trajectories of an elliptical particle in a mixture of DSPs was insensitive to the fluid types in slow flow, at higher inertial effect its lateral displacements in the Newtonian and pseudoplastic fluids differed by 0.23W.            

The discrepancy in segregation patterns and lateral displacements of the particles were more pronounced inertial microfluidics with an array of inline obstacles. Numerical simulations revealed that not only the particle shape and fluid type, but also the order of the particles at the release location had significant effect on the particles' trajectories and their separation patterns around the zone of obstacles. Although the order and orientations of non-circular particles at the release ports are difficult, if at all possible, to control in microfluidic experiments, numerical simulations revealed their non-negligible effects on segregation of particles across the inertial microfluidics. Errors in lateral displacements of particles would be as high as 0.78$W$ in these simulations, if the aforementioned assumptions are made. 

In brief, DSP-LBM simulation results demonstrated that the assumption of circular particle shape for non-circular particles and the Newtonian fluid type for non-Newtonian body fluids are not necessarily reliable and practical in assessing or optimizing microfluidic device designs for segregation of CTCs from healthy cells. Microfluidic experiments with a mixture of arbitrary-shaped particles in non-Newtonian fluids under different flow conditions can be used to test our numerical findings.

\begin{acknowledgements} 

Funding for this research was provided by Southwest Research Institute’s Internal Research and Development Program, 18R-8602 and 15R-8651. SS kindly acknowledges funding from the European Research Council under the European Union’s Horizon 2020 Framework Programme (No. FP/2014-
2020)/ERC Grant Agreement No. 739964 (COPMAT).

\end{acknowledgements}


\end{document}